\documentclass[letterpaper,twocolumn,10pt]{article}
\PassOptionsToPackage{hyphens}{url}
\usepackage{usenix-2020-09}

\def\isanonymous{0}

\def\isusenix{1}
\usepackage[l2tabu,orthodox]{nag}
\usepackage{ifxetex}

\usepackage{microtype}
\usepackage{threeparttable}
\usepackage{longtable}
\usepackage{array}
\usepackage{multirow}
\usepackage{longtable}
\usepackage{lscape}
\usepackage{xltabular}
\usepackage{tabularx}
\usepackage{siunitx}
\usepackage{hyphenat}
\usepackage{ifthen}
\newcommand{\anonymous}[2]{%
	\ifthenelse{\equal{\isanonymous}{1}}%
	{{#1}}%
	{{#2}}%
}
\newcommand{\usenixbf}[1]{%
	\ifthenelse{\equal{\isusenix}{1}}%
	{\textbf{#1}}%
	{{#1}}%
}
\usepackage{xcolor}

\usepackage{amsmath,amsfonts}  
\usepackage{xspace}
\usepackage[capitalize]{cleveref}

\usepackage[lambda,landau,operators,probability,sets,logic,complexity,asymptotics]{cryptocode}

\usepackage{booktabs}  
\usepackage{comment}
\usepackage[inline]{enumitem}
\usepackage{navigator}
\usepackage{url}
\usepackage{graphicx}
\usepackage[export]{adjustbox}
\usepackage{float}
\usepackage{multirow}
\usepackage{longtable}
\usepackage{array}
\newcolumntype{L}{>{\raggedright\arraybackslash}p{3cm}}
\usepackage{pdfpages}

\usepackage{subcaption}   
\usepackage{tikz,pgfplots,pgfplotstable}
\usepgfplotslibrary{statistics}
\usetikzlibrary{calc}
\usetikzlibrary{arrows}
\usetikzlibrary{positioning}

\pgfplotsset{
	tick label style={font=\small},
	label style={font=\small},
	legend style={font=\small, cells={anchor=west}}
}

\definecolor{DarkPurple}{HTML}{332288}
\definecolor{DarkBlue}{HTML}{6699CC}
\definecolor{LightBlue}{HTML}{88CCEE}
\definecolor{DarkGreen}{HTML}{117733}
\definecolor{DarkRed}{HTML}{661100}
\definecolor{LightRed}{HTML}{CC6677}
\definecolor{LightPink}{HTML}{AA4466}
\definecolor{DarkPink}{HTML}{882255}
\definecolor{LightPurple}{HTML}{AA4499}
\definecolor{DarkBrown}{HTML}{604c38}
\definecolor{DarkTeal}{HTML}{23373b}
\definecolor{LightBrown}{HTML}{EB811B}
\definecolor{LightGreen}{HTML}{14B03D}

\usepackage{listings}
\lstdefinelanguage{Sage}[]{Python}{morekeywords={True,False,sage,cdef,cpdef,ctypedef,self},sensitive=true}
\begin{document}

\title{On the Virtues of Information Security in the UK Climate Movement}

\anonymous{%
}{%
\author{
{\rm Mikaela Brough}\\
Royal Holloway, University of London
\and
{\rm Rikke Bjerg Jensen}\\
Royal Holloway, University of London
\and
{\rm Martin R.\ Albrecht}\\
King's College London
} 
}

\clubpenalty=1
\displaywidowpenalty=1
\widowpenalty=1

\maketitle

\begin{abstract}
  We report on an ethnographic study with members of the climate movement in the United Kingdom (UK). We conducted participant observation and interviews at protests and in various activist settings. Reporting on the findings as they relate to information security, we show that members of the UK climate movement wrestled with (i) a fundamental tension between openness and secrecy; (ii) tensions between autonomy and collective interdependence in information-security decision-making; (iii) conflicting activist ideals that shape security discourses; and (iv) pressures from different social gazes -- from each other, from people outside the movement and from their adversaries. Overall, our findings shed light on the social complexities of information-security research in activist settings and provoke methodological questions about programmes that aim to design for activists.
\end{abstract}

\section{Introduction}\label{sec:introduction}

\begin{quote}
  \textit{``The other fear is that the spy cops are using exactly the same tools. Signal, etc. And so the perception of these apps as hyper-secure makes it so that if you go into an activist setting with them, you're outing yourself as a hacktivist to a possible spy cop, or people think you might be a spy cop yourself''}~(P1).
\end{quote}

\noindent P1 is a climate activist who is part of several climate advocacy groups in the United Kingdom (UK) that use a variety of tactics, including civil disobedience, political activism and awareness raising. From the rise of Extinction Rebellion (XR), a mass movement practising civil disobedience, to the shift to more online organising during COVID-19, P1 has witnessed a rapidly changing advocacy environment, and how such changes impacted the reliance on security technology.

Environmental activism, and more recently climate activism, here refers to the actions of individuals and groups who seek to change the status quo of environmental degradation and climate change~\cite{ClappertonEA2019}. In the UK, this form of activism has a long history, with hundreds of groups making up the \emph{UK climate movement}, a heterogeneous network. Almost all groups involved in this movement present their activities as non-violent, using a diversity of tactics, including strategic voting, signing petitions, staging peaceful protests and awareness raising on social media. Some groups engage in non-violent civil disobedience, e.g.~infiltrating industry events, occupying buildings (sit-ins) and blocking roads~\cite{OrazaniSMS2021}.

The general aims of the UK climate movement align with the stated priorities of all major UK political parties~\cite{carter2014politics}. The UK Parliament also enshrined net-zero emissions targets in law in 2019~\cite{barrettEnergyDemandReduction2022}. Thus, a portion of climate activism sometimes functions as a ``quasi-arm of the state''~\cite[p.~1]{HaslerInandAgainst2020}, with some activists working within and alongside local and national authorities. On the other hand, some climate activism in the UK is criminalised, with some politicians labelling civil disobedience as a public order concern, a disruption to private enterprise and a threat to state infrastructure~\cite{HaslerInandAgainst2020,gilmorePolicingUKAntifracking2020,mireanu2014criminalisation}.

Such antagonistic political statements have coincided with extensive government spending on police intelligence within civil disobedience campaigns, in combatting what has been called a ``spectre of environmental domestic extremism''~\cite[p.~491]{schlembachdomesticextremism2018}. Accusations of domestic extremism may also mean that activists are monitored by counter-terrorism units.\footnote{E.g.~climate activists have appeared in counter-terrorist training materials for \emph{Prevent} -- while lawful protest does not meet the threshold for referrals, individuals may be classed as extremist based on their views~\cite[p.~1]{nevettPreventRiseClimate2023a}.} UK law enforcement use various surveillance and suppression techniques against climate activists. Briefly, these include: widespread use of plainclothes and undercover police officers, targeting of key organisers (through pre-emptive arrest, detention or injunctions to prevent protests~\cite{ohdedarClimateChangeLitigation2021}), intelligence sharing between public and private sectors~\cite{burdon2020targeting}, and stop and search powers. UK law enforcement agencies are also known to use advanced forms of targeted digital surveillance against social movements of concern: using tools such as drones, biometric data collection, facial recognition technology, open-source social media data collection, network analysis software and technology to intercept mobile phone traffic and location data~\cite{astonsurveillance2018,DencikhintzOINT2018} (\cref{sec:rw-uk-climate-activism}). 

The tactics of law enforcement pose significant risks to the well-being of climate activists, threatening their livelihoods, freedom and sometimes their physical integrity -- creating a chilling effect~\cite{StephensGriffinNathan2021} that affects both the moderate and more radical flanks of the movement. Yet, however powerful an adversary the British State is, it does not suppress activist movements entirely.\footnote{The term \emph{the State} here refers to the unity of the legislative, the executive and the judiciary -- ``the three main institutions'' \cite[p.~3]{benwellSeperationPowers2024}. In the United States, these would be referred to as \emph{Three Branches of Government}.} For example, neither encryption nor protest is illegal in the UK and the digital surveillance of activist groups requires certain thresholds to be met beforehand~\cite{murphy2019state}. This -- a powerful, self-limiting adversary -- shapes the security terrain for climate activists in the UK\@.

\paragraph{Contributions.} An emerging body of security scholarship has focused on the security practices and perspectives of activists in different contexts (\cref{sec:rw-security-activists}), driven by a perceived need to design for activists \emph{and} the specific needs arising from types of settings activists exist within (\cref{sec:rw-design-for-activists}). Focusing on a case that has not been studied in prior information-security work -- the UK climate movement -- we explore exactly what designing for activists and activist settings entails \emph{from the ground-up}. We employ an ethnographic approach, comprising participant observation and ethnographic interviews in various settings (e.g.~protests, community centres, parks, in activist meetings, online and trainings).

We conducted our ethnographic work in 2023 at a time when changing and stricter protest laws came into effect in the UK; a flashpoint in all kinds of UK activism (\cref{sec:rw-uk-climate-activism}). We conducted a reflexive thematic analysis, involving several rounds of interpretative coding and collaborative analysis within the research team. From this, we developed a grounded understanding of how UK climate activists conceived of and experienced their activism with respect to information security. We uncover how security is woven into the social dynamics and structures of activism, by considering the context of climate activism. We thus observe a separation between designing for activists, taking into account their social relations and structures, and designing for the operational security needs of activists that do not consider such social foundations. We show how, to successfully design for this setting, understanding how activists reason about the entirety of their activism -- and not just their threat landscape -- is imperative.

We identified four main themes our analysis. First, participants spoke of the conflict between growing the movement by enabling new members to join and the need to protect themselves and other group members. Second, we show how non-hierarchical groups experienced a conflict between the communication needs of their activism and their priorities of individual autonomy, leading to hierarchical and fragmented communication practices. Third, we found that information security was used to engage in conflicts of competing activist ideals. We observed how in-group prestige gained through non-security related skills shaped group-wide approaches to information security. Fourth, our findings point to different social gazes that influence security practices and perceptions -- peer-to-peer gaze, public gaze, police gaze.\footnote{\emph{The gaze} refers to a person's or group's sense of being perceived by others, which then influences their sense of self and behaviour~\cite{vazWhoGotLook1995}.}

\section{Policing UK Climate Activism}\label{sec:rw-uk-climate-activism}

The British State's relationship to civil disobedience is reflected in the numerous legislative changes to protest laws in the UK in 2022 and 2023 -- aimed at curbing the ``guerrilla tactics''~\cite[p.~3]{NoAuthorPublic2023} of large civil disobedience groups that grew in 2019--2020, such as XR, Just Stop Oil (JSO) and Insulate Britain (IB). For example, the introduction of the \emph{Police, Crime, Sentencing, and Courts (PCSC) Act} in 2022 resulted in five key amendments to Sections 12 and 14 of the \emph{Public Order Act 1986}. The Act broadened the definitions of offences, extended police powers and imposed tougher penalties on protesters~\cite{ParpworthCAL2022}. Further amendments to the UK's \emph{Public Order Act} (effective from May 2023) strengthened police powers to deal with protests, introducing new offences relating to obstructing traffic, delaying infrastructure and tunnelling, with prison sentences ranging from six months to three years. Among the new penalties and offences arrested persons involved in protests can now also be subject to a \emph{Serious Disruption Prevention Order (SDPO)}, which imposes restrictions on their activities, e.g.~use of the Internet, communication with certain people, access to certain places and the right to protest in the UK~\cite{NoAuthorPublic2023}.

UK authorities have extensive ``lawful surveillance powers''~\cite[p.4]{murphy2019state} through the \emph{Investigatory Powers Act 2016} (IPA) -- including the bulk and targeted interception of phone calls, emails, text messages and metadata from end-to-end encrypted (E2EE) messaging platforms~\cite{boukalas2020overcoming}. In an appeal to the Investigatory Powers Public Bill Committee, Apple submitted evidence that the ``IPA allows the UKG [UK Government] to issue secret orders to attempt to force providers to break encryption by inserting backdoors into their software products''~\cite[p.~1]{InvestigatoryPowersAmendment2023}. The IPA also contains powers for targeted and bulk equipment interference (i.e.~hacking) with a warrant and requirements that telecommunications companies retain records of phone and internet use for potential access by law enforcement agencies~\cite{murphy2019state}. Covert surveillance techniques are typically only known after the fact and, thus, the full range of tactics remain unknown. In some cases, the exact techniques used to conduct secret surveillance may never be known~\cite{murphy2019state}.

A number of UK undercover policing scandals in recent years, coupled with extensive state powers, have led activists to expect that they are under surveillance. For example, the \emph{Spycops Scandal} revealed that the London Metropolitan Police and MI5 had sent 140 police officers to spy on over 1,000 mostly left-leaning political groups from 1968 to 2008~\cite{undercoverresearchportalUndercoverResearchPortal2024}. Abuses of power during this period included using deceased children as aliases, having sexual relations and bearing children with activists, and engaging in forms of entrapment~\cite{StephensGriffinNathan2021}. These abuses were formally reviewed in 2014 through the \emph{Undercover Policing Inquiry}, with Tranche One being completed in 2023, Tranche Two hearings being completed in February 2025, and Tranche Three hearings scheduled to commence in October 2025.\footnote{See~\url{https://www.spycops.co.uk}.} This scandal increased climate activists' attention to anti-state counter-surveillance and derailed the efforts of some activists through a general chilling effect~\cite{StephensGriffinNathan2021}, and produced a tightening of \emph{activist security culture} for some people~\cite{papineaumilitant2024}.

Lastly, protest is often framed as ``socially deviant''~\cite[p.~102392]{HAYES2021102392} in popular media and considered a divisive partisan issue, shaping a hostile relationship not only with the police but also the public~\cite{wiest2015framing}. The UK has also seen cases of journalists infiltrating activist groups and passing on information to the police (e.g.~\cite{GuardianJSO2024}).

\subsection{Structures of UK Climate Activism}\label{sec:rw-structures-climate-activism}

The UK climate movement comprises varied groups with different methods and structures. We outline the key configurations and roles that give shape to their (security) strategies.

Different individual activists and groups have different relationships to authority. For example, decentralised groups often adopt flat, non-hierarchical, networked and neo-anarchic structures that reject formal leadership~\cite{FotakiMariannaERGa2022}. Many of these groups practice \emph{consensus decision-making}, where proposals on an issue are made and modified until a consensus is reached\@. \emph{Affinity groups} are small, trusted groups of usually four to six people who may share similar social goals, ideologies, identities and/or skills. Affinity groups tend to coalesce around the notion of \emph{arrestability}, defined by a willingness to be arrested during an action or confrontation. However, not all activists in the UK practice non-hierarchy, with many groups following traditional \emph{hierarchical configurations}. These groups have different ranks, starting from the entry level to more senior positions, often divided between group members and staff. Participants in our study were members of different groups with both types of organisational structure.

Within non-hierarchical groups, division of labour is typical. \emph{Organisers} are people who take on more responsibility within groups, but are not in formal leadership positions -- they are the people commonly targeted by surveillance and arrest and tend to have stricter security postures\@. \emph{Police liaisons} are group members who act as a dedicated middle-person between the group and the police during a protest. The police have a similar specialist role, known as \emph{protest liaison officers}, who wear blue vests during protests \cite{gilmorePolicingUKAntifracking2020}. Another role observed in our data is that of \emph{activist trainers}, who provide skills training within groups, including \emph{digital security training}. This can be run either by dedicated digital security organisations, or by more technologically `savvy' group members \cite{ErmoshinaConcealing2022}. During protests, \emph{legal observers}, who are trained volunteers that monitor arrests and take notes on police and protester behaviour (which can later be used in court).

\section{Prior Security Work on Activists}\label{sec:related-work}

In~\cref{sec:rw-security-activists} we draw out the insights and methods of prior security scholarship on activists. In~\cref{sec:rw-design-for-activists} we establish the call to design for activists which is the stated motivation in much prior work. Background information on the security technologies referred to in \cref{sec:findings} is in \cref{sec:preliminaries-tech}.

In the information-security literature, activists are commonly referred to as a `high-risk' or `at-risk' \emph{population}, along with other groups such as, refugees, LGBTQAI+ and survivors of intimate partner abuse~\cite{SP:WMYACKMMST22}. However, the emerging body of research on activism in security-related scholarship points to the particularities -- and temporalities -- of specific activist contexts. The broader literature on social movements shows that while activists are often assumed to share a collective identity, they adopt distinct role- and category-based activist identities~\cite{Horowitz17}. This literature further distinguishes between \emph{doing} activism and \emph{being} an activist~\cite{Bobel07}, meaning those who engage in activism may not self-identify as an activist. This work further identifies a contrast between the high-risk nature of activist settings and the mundane practices that activists engage in to create security in their daily lives~\cite{DesboroughWeldesEveryday2023,EschlePeaceCamp2018,lokotBeSafeBe2018}.

\subsection{Interview Studies with Activists}\label{sec:rw-security-activists}

Many qualitative security studies on activism involve conducting interviews that are then analysed thematically. While such studies sometimes employ in-person interviews~\cite{MCQ:Albu23,Rosenbloom22,ErmoshinaCan2017,SSR:HalErmMus18}, many studies use remote interviewing via, for example, video conferencing or messaging applications~\cite{AlbrechtCollective2021,DaffallaSecurity2021,DaffallaDefensive2021,USENIX:West17} or a mix of in-person and remote interviewing~\cite{ErmoshinaConcealing2022,CHI:KKSC16,LernerPrivacy2020,SanchesUnder2020,TadicICT2016}.

In~\cite{DaffallaDefensive2021} the authors conducted remote interviews with political activists who had participated in the 2018--2019 revolution in Sudan. They showed, for example, how these activists adopted practical workarounds, such as deleting messages and hiding applications, to protect against the threats of surveillance, arrest and device seizure posed by the Sudanese Government. Also using remote interviews, the authors of~\cite{AlbrechtCollective2021} showed how collective security practices translated to specific protective mechanisms for participants of the 2019--2020 Anti-Extradition Law Amendment Bill (Anti-ELAB) protests in Hong Kong. These included distinct practices for overcoming the shortcomings of technology used for collective action and approaches to detect compromise. Through a mix of in-person and remote interviews with transgender people, the authors of~\cite{LernerPrivacy2020} demonstrated the distinct, often rooted in identity, security and privacy risks experienced by those engaged in transgender activism. To draw out the distinction between the construction of collective and individual privacy, the authors of~\cite{CS:JiaBau22} employed the same mix of interview types. In~\cite{ErmoshinaCan2017,SSR:HalErmMus18} the authors conducted both remote and in-person interviews with human-rights activists and developers of secure messaging applications. They found disparities between what developers designed for and what these activists wanted.

Many security-related studies with activists generally adopt a structured approach to interviewing, where an interview protocol structures the engagement with participants. These protocols are typically populated with questions related to security and/or technology (e.g.~\cite{AlbrechtCollective2021, DaffallaDefensive2021}). Further, in~\cite{TadicICT2016} the authors asked 40 questions to each participant, with a focus on optimisation and software development. Rosenbloom~\cite{Rosenbloom22} conducted \emph{in-situ} interviews with Black Lives Matter (BLM) activists in the United States (US). The interviews were structured around five pre-defined questions to accommodate limited protester time. Findings related predominantly to the role of social media for information gathering, yet they also highlight specific surveillance concerns raised by participants. Other structured approaches employed to study activists' security are found in~\cite{CHI:BSCU21} and~\cite{CHIEA:WadBruFie21}, where the authors used a survey to examine the security advice given to BLM protesters.

Some prior work has established the limitations of security-related research that uses interviews as the sole data-gathering method. For example, as noted in~\cite{goyedinterviewingtrust2024}, sharing often security-sensitive and personal details requires a degree of trust between the researchers and participants. This trust can be limited in interview-only studies where a researcher may only meet a participant once, sometimes mediated through a piece of technology. Remote interviews have also been shown to limit insights on more subtle security practices and needs that can be gathered, since the medium of engagement may heighten feelings of distance between participants and researchers~\cite{loughran2022reflections}. In~\cite[p. 3367]{AlbrechtCollective2021} the authors also noted that the remote nature of their interviews with Anti-ELAB protesters may have shaped participants' responses and their ability to ``speak freely''. Yet, research-design choices (and their limitations) are not confined to data collection. Decisions researchers make about their analysis directly impact how they \emph{see} and, thus, understand their data. While generally not considered in limitation sections of prior work, qualitative information-security research often adopts a descriptive analytical approach. This is, for example, the analytical approach adopted in~\cite{AlbrechtCollective2021,DaffallaDefensive2021,USENIX:SCBAPB22} to mention a few. Some authors (e.g~\cite{USENIX:SCBAPB22,USENIX:MccJenTal23}) have also called for a diversification of research designs in information-security research to overcome some of the limitations of interview studies.

\paragraph{Engaging Activists \emph{in the Field}.}\label{sec:prior-work-field} A few studies exploring activism and security have engaged with activists in different field settings, employing both \emph{in-situ} interviews and participant observation. The authors of~\cite{AsadIllegitimate2015} conducted fieldwork with local housing justice activists in various settings in the US. Focusing on a four-hour window during a protest, they showed how activists used digital technology to draw attention to their cause and garner support. The authors of~\cite{WulfFighting2013} conducted a mix of fieldwork and digital ethnography to understand the role of social media among Palestinian activists in 2013.\footnote{Digital ethnography has been used in some security research on activism~\cite{lokotBeSafeBe2018,TadicICT2016}; yet, this method is limited to online observations (e.g., exploring social media posts and chatting online).} Focusing on tensions between visibility and security among Moroccan activists, Albu~\cite{MCQ:Albu23} used a mix of interviews and observations in different Moroccan settings, including at demonstrations. They showed how encryption, obfuscation and concealment, at times, both empowered and undermined the efforts of activists and their adversaries. In their study, the authors of~\cite{ErmoshinaConcealing2022} drew on Science and Technology Studies (STS) to examine how encryption is experienced differently by those who rely on it and those who develop it.

\subsection{Technology Design and Activism}\label{sec:rw-design-for-activists}

A key driver of much security scholarship on activism is to \emph{design for activists}. This is evident from how authors motivate and discuss the significance of their work by, for example, setting out a series of design recommendations. In~\cite{AlbrechtCollective2021} the authors, albeit cautiously, set out some recommendations for secure messaging, including the sharing of a live location securely and improved group administration to manage membership. In~\cite[p. 760]{SanchesUnder2020} the authors noted that to ``design for the dissident'' one needs to design for plausible deniability and the ability to control facets of identity. Hirsch~\cite[p. 10]{MIT:Hirsch08} advocated ``contestational design'' to design for different political activist contexts, e.g.~Dialup Radio for Zimbabwean activists~\cite{CT:Hirsch09}. In~\cite[p. 8]{AsadIllegitimate2015} the authors avoided prescribed solutions, but proposed a generalised design approach of ``flexibility and process''. Other security studies on activists provide advice on what activists should do, e.g.~sanitise phones before protests, use Multi-Factor Authentication (MFA), secure messaging, etc.~\cite{DaffallaDefensive2021,DaffallaSecurity2021}.

Several studies that aim to serve activists also point to specific collective ideas of protection. In~\cite{SanchesUnder2020}, for example, the authors showed how the participants in their study felt a sense of individual futility in terms of being under surveillance and sought security communally; thus arguing against an individualist framework of activist protection. Similarly, the authors of~\cite[p. 3374]{AlbrechtCollective2021} showed how Anti-ELAB protesters experienced a sense of ``security in numbers'' and how they employed protective practices to fulfil their own security needs \emph{as well as} those of the group. Other work has foregrounded the political aspects of design choices, with the authors of~\cite{ErmoshinaConcealing2022} drawing attention to schisms in understanding. This builds on prior work that identified a disconnect between what developers of secure messaging design for and the needs of human-rights activists~\cite{ErmoshinaCan2017}. The authors stressed that to appropriately prioritise design properties, more dialogue between designers and those being designed for is integral. In~\cite{FCJ:AGRS15} the authors advocated for more grounded approaches to technology development to enable technologists to design for activist needs.

In thinking about what \emph{designing for} means, the authors of~\cite[p. 3883]{CHI:KKSC16} use the idea of ``undercurrents''; invisible activities that happen on the ground that frame decision-making in activism. To them, security practice is the observable output of these undercurrents, and designers must understand these foundations to contextualise technology decisions. Our work directly responds to this call within security by bringing attention to the security undercurrents of the UK climate movement.

\section{Research Design}\label{sec:research-design}

Our research design adheres to ethnography; an approach that constitutes
\begin{enumerate*}[label=(\alph*)]
\item theoretical \emph{ways of seeing} (e.g.~through relational social interdependency),
\item methodological \emph{ways of doing} (e.g.~participant observation) and
\item practical \emph{ways of engaging} (e.g.~through social embeddedness)~\cite{BrewerEthnographyBook2000}.
\end{enumerate*}

\subsection{Methods}\label{sec:methods}

Our study involved participant observation in climate activist settings in the UK and ethnographic interviews with members of the UK climate movement. Through the combination of interview and field data, ethnographic approaches can bridge the gap between \emph{what people say} and \emph{what people do}, while also considering their social contexts.

\paragraph{Participant Observation.}\label{sec:participant-observation} Participant observation was conducted at meetings, street protests and training sessions, with the researcher participating in the interactions of the groups under study~\cite{BruynPOBook1966}. Access to these settings was granted through key interlocutors, while the researcher's presence in these settings was fully transparent (\cref{sec:ethical-considerations}). In line with established ethnographic practice~\cite{GeertzIOC1973}, full field notes were taken during or immediately after an observation in the eight field settings (\cref{tab:participantobservation}). Full field notes included the researcher's observations of events and conversations, coupled with reflections on them with possible interpretations. Such field notes provided key contextual details for analysis~\cite{PhillippiGuide2018}. Observation took place from May to July 2023.

\begin{table}[h!]
	\caption{Participant observation sites.}\label{tab:participantobservation}
	{\footnotesize\centering
		\begin{tabular}{lrrlrr}
			\toprule
			Setting Type & T  & P      & Setting Type & T & P \\
			\midrule
			Protest 1    & 3  & 75     & Meeting 2    & 5 & 40 \\
			Protest 2    & 16 & 60,000 & Meeting 3    & 5 & 30 \\
			Protest 3    & 8  & 150    & Meeting 4    & 3 & 30 \\
			Meeting 1    & 1  & 250    & Training 1   & 3 & 25 \\
			\bottomrule
		\end{tabular}\par}
	\vspace{\baselineskip}
	{\footnotesize ``T'': rough time in hours, ``P'': approximate number of people.\par}
\end{table}

\paragraph{Ethnographic Interviews.}\label{sec:interviews} 15 ethnographic interviews were conducted, nine of which were conducted \emph{in situ} (\cref{tab:participants}), and were loosely structured around an interview guide (\cref{app:appendix-a}). Interviews took place from June to July 2023. Participants were recruited using convenience and snowball sampling~\cite{GuestHandbook2014}, sometimes as a result of contact made through participant observation. Each participant was asked if they wished to be audio-recorded, with 10 participants preferring not to be recorded. Recorded interviews were later transcribed and unrecorded interviews were captured through hand-written notes during the interview and expanded into detailed descriptions the same day, following the guidelines provided in~\cite{RutakumwaConducting2020}. 13 participants had taken part in civil disobedience, among other methods. To protect participants' identities, IDs were randomly assigned. We also assigned a letter to each group so that participants in the same group can be identified through the data (\cref{tab:participants}).

While interviewing is a typical qualitative research method, ethnographic interviewing is a distinct form of this method. Generally, social research looks to interpret a given field through both declarative (what is said) and non-declarative (what is unsaid) knowledge. In ethnography, the interview is not only a process of declarative knowledge access, but also a non-declarative social experience~\cite{rinaldoguhin2022interviews}. This style of interviewing, coupled with participant observation, enables rapport-building over time, is generally marked by a higher degree of trust and informality between researchers and participants than more conventional interviewing~\cite{rinaldoguhin2022interviews}. Ethnographic interviews are part of wider field relations and field data; they are shaped by observations made by the fieldworker \emph{in situ}, allowing for situational insights and questions to shape the direction of interviews. This leads to an evolution of topics -- sometimes spontaneous -- throughout the fieldwork and is responsive to developments over time. 

Through the interviews we collected both declarative and non-declarative knowledge, which we triangulated with non-declarative knowledge from participant observation. As mentioned, not all interviews were conducted \emph{in situ}, as some were conducted remotely for the convenience and comfort of some participants. These participants were, however, also involved through field engagements such as observations at group meetings which allowed for social contextualisation of all interview findings.

\begin{table}[t]
	\caption{Interviewed participants and group membership.}\label{tab:participants}
	{\scriptsize\centering
		\begin{tabular}{lclr @{\hskip 2em} lclr}
			\toprule
			ID  & G & Mode      & T & ID  & G & Mode      & T\\
			\midrule
			P1  & A,E,G & Online   &  90 & P9  & B     & In-Person & 120\\
			P2  & F     & Online   &  60 & P10 & J     & In-Person &  30\\
			P3  & E     & Online   &  60 & P11 & E     & Online   &  90\\
			P4  & K     & Online   &  90 & P12 & C,D   & In-Person & 120\\
			P5  & H,I   & Online    &  90 & P13 & J     & In-Person & 120\\
			P6  & B,C,K & In-Person &  60 & P14 & K     & In-Person & 120\\
			P7  & A,B   & In-Person &  90 & P15 & D     & In-Person & 120\\
			P8  & B,K   & In-Person &  60 & \\
			\bottomrule
		\end{tabular}
		\par}
	\vspace{\baselineskip}
	{\footnotesize Some participants were part of multiple groups. Typically, multiple group membership involved groups of different sizes that used different tactics. ``T'': rough time in minutes,  ``G'': groups. \par}
\end{table}

\subsection{Reflexive Thematic Analysis}\label{sec:data-analysis}

We followed Braun and Clarke's~\cite{BraClar19} approach of reflexive thematic analysis, which suggests that text gains significance through the iterative construction of interpretive codes, categories and code groupings that organise underlying meanings within the text~\cite{BraunOne2021}. This analysis specifically encourages us to question the subjective foundations of our decisions, prioritising inductive reasoning~\cite{BraunandClarkeTA2022,BraunOne2021}.

The researcher who conducted the observations and interviews (hereafter: the \emph{fieldworker}) first analysed the transcripts, interview descriptions and field notes by making descriptive and interpretative summaries of the complete data set, after this had been segmented into smaller analytic units. The research team then engaged in an interpretative round of summaries, allowing for alternate interpretations of existing summaries and explorations into how the researchers perspectives might influence our understandings of the data. This was followed by \emph{open coding}, where concise statements (codes) were constructed for each distinct idea observed during the entire summarising stage. Two coders (including the fieldworker) began this phase independently, each generating initial codes representing distinct ideas from the summarised material. These codes were then reviewed and consolidated through collaborative coding sessions involving iterative comparison, discussion, and refinement. Through this process, codes were revised, merged, or reorganised around emergent patterns. This resulted in 29 codes, which were then axially grouped into themes. The collaborative analysis continued through the writing phase, which is considered one of the analytical steps in thematic analysis~\cite{BraCla06}. This process led to the four themes that structure our findings in~\cref{sec:findings}. 

The 15 interviews conducted were not based on a pre-determined sample size, but materialised organically through the course of fieldwork, consistent with the adaptive nature of ethnographic research. During the coding and theme development phases, we reached a point where further coding and re-coding of the data no longer yielded new codes. Rather than a numerical endpoint, this form of `thematic saturation' was interpretative; at this stage of the analysis the themes were of sufficient depth, coherence and analytic richness to underpin our findings as no new insights were materialising.

\paragraph{Researcher Positionalities.}\label{sec:res-pos} A distinctive feature of ethnography is reflexivity, which acknowledges the researchers' potential impact on the settings and groups they study, while analysing their own subjectivity~\cite{DaviesReflexivity1999}. Our individual and collective identities shaped the data and how we interpreted it. The fieldworker is a non-UK born woman (she/her) and trained ethnographer who, although not involved in climate activism, has experience in other forms of advocacy. The research team also includes another ethnographer who researches information security as it relates to people at the margins of societies and a cryptographer whose interest in the field grew out of activism and the resulting experiences of surveillance. No author identifies as a climate activist.

\subsection{Limitations}\label{sec:limitations} 

Several limitations shaped our study. Participant observation took place over short periods of time at protests, meetings and trainings. The use of snowball and convenience sampling to recruit participants meant that those interviewed were also loosely connected. However, several interviewees were recruited at meetings and protests, which reduced the effects of self-selection. There is always a risk in using interviews to study security, since politically charged questions may skew the interview sample towards those who feel comfortable discussing such issues or have strong opinions. However, we sought to overcome this limitation by adopting an ethnographic approach, while adopting a reflexive analytical approach.

\section{Findings}\label{sec:findings}

We organise our findings in line with the themes created through the reflexive thematic analysis (\cref{sec:data-analysis}). In~\cref{sec:dilemma}, we outline how participants constructed ideas about adversaries and their capabilities, and how security considerations shaped group social formations. In~\cref{sec:necessity-freedom}, we highlight the tensions between the observed necessities of climate activism and ideas of individual autonomy, through a focus on secure messaging application choice. In~\cref{sec:findings-power-dynamics}, we show how dynamics around information security within groups derived from distinct activist ideals and virtues. \Cref{sec:findings-social-gazes} presents findings on how different social gazes -- peer-to-peer gaze, public gaze and police gaze -- shaped participants' activism and, thus, information security.

\subsection{Fundamental Security Tensions}\label{sec:dilemma}

Our analysis revealed distinct tensions within the groups under study, grounded in conflicting views on security.

\subsubsection{Constructions of Adversaries}\label{sec:adversaries}

Our data suggests that there is both a temporal and subjective view of who or what constitutes an adversary for UK climate activists. These multiple views are shaped by intertwined experiences, stories, subcultural preferences and histories, and are constantly made and remade. However, the most common and concrete immediate adversary discussed by participants was law enforcement, with the London Metropolitan Police (often called `The Met') being considered by some participants as the most antagonistic police force in the country in policing protesters. Law enforcement was seen to undermine the effectiveness of a given protest either by breaking it up before it started or by breaking it up as it happened. 

Participants highlighted how they would try to protect group members and information through practices whereby compromised parties could be sanitised from group chats, for example. This was framed as a way to avoid exposing future group plans or other group members in the event of device confiscation. P11 also recounted instances where some UK climate activists had been followed by the police: \textit{``All the way home from the protest in London, Group C was very obviously followed by police''}. P11 noted that this type of physical surveillance had stoked fears within their group and acted as a reminder that they were being \textit{``watched''}. P11 shared how stories and experiences of police violence against protesters \textit{``leaches into a latent fear or worry''}, further highlighting concerns about \textit{``police trying to infiltrate online spaces''} stressing that \textit{``there is a need for secure chats in terms of that''}. Thus, experiences of on-the-ground police surveillance had fuelled fears of online surveillance for P11.

Not every participant considered the police adversarial. Conflicting views on law enforcement were observed within the same groups. P9 noted that it was a \textit{``matter of perception''}, highlighting that people who \textit{``believe that there are forces that monitor and control us''} would find it necessary to secure their communications, whereas those who \textit{``believe in what the government is saying publicly''} would not. P9 explained that both perspectives were present within some groups and it was noted that this `matter of perception' was also influenced by how UK climate activists perceived the righteousness and legitimacy of their own tactics. P6 noted that those who saw their strategies as \textit{``palatable''} tended to fear the police less, while groups characterised as disruptive or unpopular with the public tended to worry about police surveillance more.

The necessity of security was shown to be subjective and differed depending on the time and place. For example, security was observed as most important to activists during and immediately before protests and acts of civil disobedience, when the risk of police infiltration was perceived to be highest. Communication in the days leading up to a protest required secrecy to prevent the protest from being pre-emptively stopped. Communication during the protest also needed to be secure so that the police could not use phones to intercept communications on the spot. Participants reported sometimes leaving their phones at home for this reason and espoused this view at meetings. The days after a protest were also observed as critical to ensure that confiscated devices could be dealt with, restoring the group's secure channels.

\subsubsection{Security Conflicting with Movement Building}\label{sec:secMB}

Participants generally argued that security and movement building were at odds with each other. Security was seen as increasing the complexity of a group's technological infrastructure and raising \textit{``the bar for entry''} (P2). P2 explained:

\begin{quote}
\textit{``And so local groups will plan actions and some of them will be various levels of risk and stuff. And the tension between using a secure and encrypted messaging platform \dots but then like, does that turn away potential new members? Or make it a high bar for entry? I feel we're always wrestling with that question in different ways. And I guess a typical thing is that we use different platforms for different things. But then this becomes really complex.''}
\end{quote}

\noindent Ideas of security were often described as embodied in practices such as the use of encrypted messaging platforms, but the use of such tools was also seen as potentially discouraging new people from joining the movement. As P2's group used different platforms for different organising tasks,\footnote{The group used Action Network to advertise protests, CiviCRM to manage donations and Signal to create affinity groups.} there was a collective desire to minimise the number of new tools that members had to learn. Perceptions of accessibility were shaped by an understanding of the tools with which members were already familiar and, thus, comfortable. In addition, it was seen as important for the growth of the movement that the groups appeared mainstream and \textit{``normal''} (P2) to anyone outside the movement. Secure technology was generally seen as working against this goal, as it was perceived as both intimidating and reinforcing the idea that the groups were \textit{``too fringe, or like too out there to join''}  (P3).

\subsubsection{Creating Informal Boundaries as Security Practice} 
\label{subsec:boundaries}

The tension between growing the movement and maintaining strict(er) boundaries coexisted within many groups. Despite the stated goals of inclusivity, we observed a process of boundary making that served to protect a group from infiltration. Some participants noted a division of trust within their group, where public-facing groups (e.g.~broadcast channels, mailing lists) did not contain integral information. At the other end of the spectrum, small affinity groups centred around private group chats, often on a platform chosen collectively by the group (e.g.~Signal, WhatsApp, Threema, Telegram, Session).

We observed that the trust assumption for the public-facing components of activism was simple: everyone was assumed to be malicious by default. Conversely, we also observed that the trust assumption for affinity groups was simple in a different way: all members were assumed to be honest actors. It is at the edges, or blurring, of these simple assumptions that we see more sociological processes of \emph{boundary making}\footnote{Boundary making (related to boundary work) is a well-accepted sociological concept referring to socially constructed ways that groups differentiate each other, expressed symbolically through cultural attitudes, practices, and taboos~\cite{Zurbaboundaries2022}.} structuring ideas of trust vs.~distrust. In other words, individual judgements about what was in-group vs.~out-group behaviour were formative for the groups. In \emph{Meeting 3}, some participants described this as \textit{``acting weird''}, where perhaps somebody was seen to be fidgeting, saying something that was perceived as contradictory, or being agitated by certain types of political rhetoric used by group members. Assessments of in-group and out-group behaviours underline the tension between movement building and security, as group members outlined norms of behaviour for UK climate activists.

Our analysis showed that while affinity groups were typically made up of people who all knew each other personally, participants gave examples of how the boundaries of such groups were sometimes blurred. P13 explained that some people in their small hometown did not have the opportunity to connect with their group in person. This made them feel alienated from the group because they felt that the in-person core of the group formed a cloister of trust. Participants generally explained that they found it difficult to assess whether a new member was trustworthy (and not a secret police officer, someone from a right-wing group or a tabloid) using only technological means. Affinity groups were therefore formed in person, either well in advance of a protest or, in situations where people had to travel to the protest site, just before a protest. Last-minute affinity group formation tended to rely on personal cues and alignment with core values (e.g.~\emph{arrestability}; see \cref{sec:rw-structures-climate-activism}). Participants argued that it would be harder for people to disguise dishonesty in person, as being in the same physical space was thought to help detect deceit through the aforementioned markers of \textit{``acting weird''}.

P2 recounted a time when they were trying to organise a \textit{``high-risk action''} involving a lot of people they had not met in person: \textit{``Sometimes we were like `this guy seems super sus or whatever'. But I think in the end there's not that much you can do anyway because you cannot just be like `oh, this person is really sus' -- what are you going to do? Ask them?''}
Using words such as \textit{``weird''} and \textit{``sus''} (slang for suspicious) about potentially untrustworthy persons, participants revealed boundary making and individual safety assessments. 

\subsection{Autonomy and Dependency}\label{sec:necessity-freedom}

Some groups in our study adopted non-hierarchical organising principles; that is, no formal relations of authority, consensus-based decision-making and a strong emphasis on autonomy. The purpose of any such group existed only through the collective will of its members, and there was no other entity -- whether personal (e.g.~a leader) or impersonal (e.g.~a committee) -- that represented this will of the group. Yet, our findings suggest significant tensions between the agreed necessities of their activism, on the one hand, and their autonomy and particularity, on the other. We use data from Group~E to explore this tension.

\subsubsection{The Case of Group E}\label{sec:group-e}

Group E is one of the non-hierarchical, consensus-based groups in our study. It had a medium-sized core (P1, P3 and P11 are members) and a diverse secure messaging ecosystem. P11 reported that Group E had exclusively used WhatsApp until certain group members started advocating the use of Signal in 2021: \textit{``I don't really know why Signal is more secure than WhatsApp, but I think it's probably to do with Meta, at least that's what the guys said''}. Although the group had a meeting to decide to switch to Signal, Group E did not fully adopt the switch from WhatsApp to Signal. P3 noted that they also used WhatsApp for their activism, since they found security \textit{``very difficult''} and \textit{``made everything less functional''}. 

We found that group members with an interest in technology (hereafter referred to as ``digitally-inclined'' (P11)) provided advice on which technologies the group should use. It was argued by P1 that since activists generally saw technology as an attack vector for law enforcement, much of this advice prioritised specific counter-surveillance properties. Since group members were ultimately free to use whichever applications they wanted (also reflecting priorities beyond security), there was a tension between what digitally-inclined group members proposed in collective-decision meetings and what other group members wanted for themselves: \textit{``It's a weird thing when you have a group of people who are so hyper secure and then you have people [in the same group] who are so low security. It's an interesting contrast''} (P1).

Partially due to the segregation created by the use of both Signal and WhatsApp within Group E and partially citing the usefulness of channels for separating working groups from each other, P11 described how the group eventually decided to migrate to Slack.\footnote{On Slack, P2 also noted: \textit{``Slack is a specific kind of corporate online workplace culture vibe. And so if you're trying to build a movement of like working class people, a lot of them are like, what the hell is this app you're making me download?''}} Despite this, many Group E members did not move to Slack, but chose to stay on WhatsApp or Signal. Therefore, Group E considered technical perspectives on communication and collectively decided to heed this advice. Despite this, they did not act upon this decision collectively. P11 underscored this by noting that the practices were \textit{``a lot less formal than people realise''}.

This disconnect -- between collectively agreed upon security practices and what people actually did -- was also highlighted broadly beyond secure messaging application choice. During protests, we observed that some \emph{actual} information-security practices did not reflect some of the \emph{described} practices. For example, phones were frequently \emph{not} left at home because important information was constantly being exchanged on a variety of applications and in different groups. While some people had burner phones, these phones were sometimes not in important group chats that contained necessary updates on locations and arrests. Since usually only dedicated police liaisons (\cref{sec:rw-structures-climate-activism}) spoke with police officers, the outcome of these interactions (e.g.~relevant warnings of arrests) were considered essential updates for other protesters. Therefore, there were barriers to organisers' efforts in \textit{``ramp[ing] up''} (P13) security when necessary (e.g.~by telling protesters to leave phones at home).

\subsubsection{The Paradox of Autonomy}\label{sec:paradox-autonomy}

The tensions in decision-making described by members of Group E thus led to a situation where, on the one hand, common communication was generally accepted as a necessary precondition to achieve collectively-agreed goals, and where, on the other hand, incompatible standpoints of autonomy made this common communication impossible at the technology level. This contradiction did not reach a crisis point (i.e.~a breakdown of communication) because group members were using several communication channels and organisers translated between them. That is, participants experienced challenges such as \emph{cross-posting} (organisers disseminating identical logistical information across multiple applications) and \emph{message oversaturation} (for people who had every application). These cross-posting efforts were described as exhausting for the people responsible for disseminating crucial information. The observed dynamic -- autonomous standpoints reject the necessity of common communication -- did not alleviate the necessity but introduced a new, additional interpersonal dependence: on organisers to disseminate information by cross-posting. Simply put, the insistence on independence when faced with necessities introduced new dependencies.

This practice also proved exhausting for those receiving a lot of information: \textit{``So it's like, you have 10 different Signal chats and also there's like a Slack and then also there's like an email list and then it's just overwhelming for most people on everything''} (P2). Some participants highlighted how they chose to mentally disengage from their online chats. For example, P12 explained they felt that messages became redundant because they were repeated across applications and chats. The message oversaturation experienced by participants referred both to the form (which applications they chose to engage with) and the quantity (too many messages being shared). Many participants noted that they preferred to have just one application dedicated to their activism, but that this was not possible due to the heterogeneous technology adoption within groups. For example, P3 preferred to use only Signal for activism, WhatsApp for family and friends and Slack for work. However, as Group E cross-posted between all these applications, P3 said that they would try to mute all chats outside of Signal. In this way, P3 felt they could compartmentalise their life and control information flows.

\subsubsection{Deliberate Fragmentation}\label{sec:deliberate-fragmentation}

Some groups intentionally fragmented their communications along lines of risk, engaging in cross-posting for security. For example, members of Group B explained that riskier actions would be organised on Signal, whereas Telegram was described to be used for more public-facing communication. Telegram and Signal played a significant role in organising \emph{Protest 2}. Pickets, speeches and action strategy were organised in advance using these two applications. For example, all legal observers were connected both on Telegram and Signal, with the Telegram chats being larger and the Signal chats being for smaller groups, discussing more sensitive information. At \emph{Protest 2}, participants and organisers constantly complained that in spite of security, the need to cross-post was frustrating and took away from the protest experience. As P7, who was an organiser, put it: \textit{``I hate technology, I always forget where certain information is living [in either Signal or Telegram] at actions \dots most of the time we argue in the chats about being able to find certain information.''} P7's frustrations were related to the functional inability to seamlessly copy and paste key information across Telegram and Signal. Hence, some organisers abandoned the use of their phones during protests, so certain essential information (e.g.~maps or times) was not shared when needed. P5 highlighted the confusion arising from this: \textit{``I always forget which one I even use, right now there's like four or five. There are things on different apps. You might have like, your own subgroup you make on WhatsApp that comes out of Slack or something.''}

In such intentionally fragmented environments, dividing activities across applications derived from a desire to mitigate surveillance. P8 explained that if law enforcement seized a device, Group B's activities would be spread across multiple applications and many chats. P8 argued that this would make it harder for the police to gather evidence from a single application or chat: \textit{``We don't want to put all our eggs in one basket.''} Further, some participants also highlighted the use of code names as a security feature. Activist trainers at \emph{Training 1} argued that the use of only code names instead of real-life names in messaging (and using a different code name for each application) added an \textit{``extra layer of resistance''} if they were arrested. The logic underpinning this advice was that it would make the police's job of determining who is who between applications more difficult. By leveraging the reality of a diverse ecosystem of applications in this way, fragmentation was praised as a means of strengthening both confidentiality and deniability by making it harder for law enforcement to re-combine and connect information across applications. This advice was given both in cases where fragmentation could not be avoided (in groups that disagreed on application choice) and where it was mandated (in line with threat perception).

\subsection{Tensions Between Activist Ideals}\label{sec:findings-power-dynamics}

Through the reflexive analysis, different social dynamics became visible as the data highlighted how the ``ideal activist identity''~\cite{CraddockIdealActivist2019} underpinned some observations that participants had about their own security -- and, prominently, the security of others. We capture these observations by showing how participants answered questions about their own practices and competences by speaking about others; demonstrating that, for the groups under study, security served as a framing for broader negotiations of personal and collective identity.

\subsubsection{The Self through Others}\label{sec:findings-power-dynamics-ACEO}

When asked about their own security perceptions and practices, participants often gave their views on others; other people within their own groups and/or in the wider movement. Thus, while participants sometimes claimed that they did not have the necessary expertise in security technology to give an informed opinion or gave a general self-assessment such as \textit{``I am a bit intimidated by it''} (P7), they would sometimes feel able to comment on other people's security practices. For example, when asked whether they considered security to be a priority, P5 responded: \textit{``I think it's not something I directly interact with.''} Despite this, P5 also noted that other people in the movement \textit{``are mostly anti-technology, but that's not my perspective \dots I have heard stories of people going into meetings and putting their phones in microwaves. I do not believe in making change this way.''} P5 generally perceived other group members' efforts to resist online tracking through the use of encryption and private search engines as futile. Through this reflection on other people's practices, P5 seemed to self-define as someone who felt a low sense of agency when it came to their security and privacy, while expressing a low desire for the project of resisting online tracking.

A similar position was evident from P1's interview where they, rather than discussing their own practices, referred to the practices and security beliefs of others: \textit{``It's weird because some people do not want a digital footprint, like they take cybersecurity to the absolute max.''} However, P1 also noted: \textit{``A lot of people have this laissez-faire attitude to security until it's too late.''} By assessing the practices of others, P1 assumed a security position somewhere between these two approaches -- without stating or reflecting on their own personal practices in the process. In a similar vein, P14 claimed that practising stringent information security could position activists on the fringe of groups, since \textit{``it is not a top priority for most''}. With this assessment being about others and not themself (they did not refer to themself as being on the fringe), P14 may not have prioritised security as a core aspect of their activism.

\subsubsection{The Intersection of Ability and Morals}\label{sec:techbros} 

Several participants commented on how they perceived other group members' technical competencies. These assessments came from members who would be considered digitally inclined. P9 explained: \textit{``People are not actively malicious, but people do things out of ignorance and when you are the person who always has to pick up the pieces, it drives you crazy.''} P2 shared a similar perspective, although their assessment was influenced by age-specific assumptions: \textit{``We used to have quite a few older people and I think that was a challenge \ldots now we have lots of students and they teach us things which is really cool \ldots it makes it kind of hard sometimes too.''}

P11 gave the example that many younger members had wanted to move away from using Zoom because of concerns that the platform provided facial biometric data to law enforcement, raising this as unethical.\footnote{Little is publicly available about law enforcement's sources of biometric data used for facial recognition technology, although the use of facial recognition is widespread at protests in the UK~\cite{fussey2020policing}.} Yet, Group E also comprised \textit{``older members''} (P11) who they thought would struggle with a new tool: \textit{``It's a big part of why we do everything on Zoom, for those who aren't as tech-savvy because of their age''} (P11). Considerations here were thus based on assumptions influenced by dominant societal discourses on technological literacy, such as older people struggling to use technology would trump the ethical concerns of a certain platform.

While P11 framed this through a benevolent rhetoric, they uncovered the strong ethical imperative that some members in P11's group felt about switching away from Zoom. That is, rather than framing the provision of facial biometric data to law enforcement agencies -- the main adversary considered in this context (\cref{sec:adversaries}) -- as an immediate security concern, this provision was framed as a moral question, broadly related to corporate tracking. We thus found that while groups often praised technological accessibility, assessments of group members' technical competence were sometimes rooted in considerations of virtue. This meant that rather than framing technical knowledge gaps as security concerns, they were framed as moral questions.

The assessment of the technical competencies of others was also connected to a broader critiquing of practices that did not conform to what was considered typical activist practice, and the types of values that are laden in this norm. Our data shows how the idea of an ideal activist identity was constructed through education, social positioning and ``doing enough of the right thing''~\cite{CraddockIdealActivist2019}. For example, P6 noted that security decisions were sometimes influenced by individuals who advocated for a particular cause in solidarity with global efforts against surveillance. P9 explained that they would make decisions about their technology use in light of other core values related to anti-surveillance. This included advocating the use of open-source software within their group even if this was not strictly related to security: \textit{``It would be a public good if we used open source software for security and beyond''}. P1 noted that some groups (e.g.~Group J) had chosen Session over Signal because \textit{``they don't collect user data, they're open source, they don't have to use a phone number, they use strong encryption''}. While P1 outlined security reasons for the adoption of Signal, P5 explained how decentralised applications could be adopted also for ideological reasons, since they \textit{``model the ideologies of the protesters \ldots creating these autonomous zones.''}

\subsubsection{Security as Paranoia}\label{sec:luddites}

While digitally-inclined participants commented on people who they considered to have a (too) relaxed attitude to security, the latter group often criticised those considered digitally inclined for being paranoid. In so doing, they demonstrated the contradictory construction of ideal activist identities, in relation to information security. In particular, the more anarchist factions within the movement were often described as \textit{``overly paranoid''} (P1), which was seen to hinder their ability to engage in meaningful collaboration with other members of the movement. As P2 put it: \textit{``It can be like a kind of paranoia that comes with secure people. That's like, not a good vibe.''} At \emph{Protest 3}, one protester noted: \textit{``I know a bit about security, mostly because I know a lot of paranoid people.''} People who did not consider security technology to be important within the climate movement tended to judge those who cared about security issues, deeming them to be exclusionary and intolerant to other perspectives. In other words, information security was seen as discriminatory and off-putting to people from certain demographic groups, with P3 stating: \textit{``I think accessible in the sense that it is intersectional, I think a lot of groups are trying to make sure that inclusivity is practised.''} Thus, certain security approaches adopted by some activists were considered paranoid and, by extension, unethical, because they were seen to violate what some participants saw as a strong norm of inclusion within the movement.

This divide between \textit{``paranoid''} and \textit{``inclusive''} meant that people with similar views on security often gravitated towards each other. For example, P9 highlighted their strong preference for Signal over WhatsApp, citing ethical concerns about privacy policies. When asked if this choice hindered their communication within the climate movement, P9 explained that all their friends were Signal users. P9 also expressed a reluctance to befriend people who used WhatsApp. P5 noted: \textit{``Because it's a bit of a herd mentality \ldots If you have like 10 people at a time who have super secure views, they are basically trying to connect with other 10 people who have super similar views.''} While the desire to demonstrate a preference of some applications over others, e.g.~Signal over WhatsApp, highlights distinct technological preferences it also signals particular virtues, e.g.~anti-surveillance.

Overall, the prioritisation of subcultural preferences in conjunction with operational concerns indicates that doing activism involves normative considerations of how people think of themselves and others around them. Security exists here as a framing for this identity-related dilemma.

\subsubsection{Activist Prestige Framing Security}\label{sec:prestige}

Our data shows that participants often constructed their security practices and perceptions based on members they saw as possessing significant social standing within the movement.

The accumulation of social prestige in activist circles was observed in our data to occur through the acquisition and demonstration of an educated and practically skilled \emph{habitus} (personal habits, skills and dispositions)~\cite{BourdieuDistinction2010}. For example, P12 expressed their admiration for an activist within their group, praising this person's high self-esteem and competence as an activist. They admired them for participating in protests despite changes in protest laws in England, expressing a desire to feel equally \textit{``confident''}. Valued knowledge included environmental science, social movement strategy, campaigning and race/gender theory. Certain lifestyle traits were also observed to contribute to the accumulation of social prestige, e.g.~having an affinity for outdoor activities.

Our data contains examples of how members who possessed a significant standing in the group would provide advice on security, even if their social prestige had been accumulated by demonstrating a non-security knowledgeable skill, practical ability or the projection of a counter-cultural image. At \emph{Meeting 1}, an organiser with a significant social standing -- but not one of the digitally inclined -- espoused the use of Telegram for secure communication. This led attendees at this meeting to download the application even if some expressed a reluctance to do so. We thus observed that activists with high in-group prestige sometimes were mirrored by others. This suggests that ideas of self within the UK climate movement sometimes derived from evaluations of others.

\subsection{Pressures of Different Social Gazes}\label{sec:findings-social-gazes}

How participants approached their activism was further shaped by intertwining social gazes~\cite{vazWhoGotLook1995}. By framing their security perceptions through the perspective of others, participants expressed concerns about potential negative judgements from their social circles, constituting a reciprocal social gaze. This fear was influenced by three different perspectives: the \emph{peer-to-peer gaze}, the \emph{public gaze} and the \emph{police gaze}.

First, our data uncovered how participants experienced a considerable pressure to conform to socially accepted practices and to embody an ideal activist identity in relation to security (\cref{sec:findings-power-dynamics-ACEO}). For many, using some amount of encryption was seen as a necessary step to avoid the judgment of others. For example, P7 stated that they preferred to use encrypted messaging applications such as Telegram and Signal because it ``\emph{makes me feel like I am doing something right}''.\footnote{Many participants thought that Telegram offered E2EE by default. Yet, E2EE in Telegram must be activated in a `secret chat' between two parties, and is not offered in group chats.} Some participants expressed a sense of shame for not being more knowledgeable about information security. During \emph{Protest 2}, several protesters discussed how they felt that they needed to improve their understanding of security technology, admitting that it was one of their ``\emph{blind spots''}. Yet, participants also noted that they would make a conscious effort not to appear more or less paranoid than their peers to avoid standing out. P6 was concerned that having secure messaging applications other than WhatsApp on their phone might mark them as a hacktivist: \textit{``People who care most about having apps other than WhatsApp seem to do so out of a personal interest in technology, or they are hacktivists.''}

As observed in meetings, there was a tendency among participants to wonder if people who were interested in technology were potentially undercover police. This observation came to light during periods of manual group chat formation and membership amendment. P1 highlighted how both police and hacktivists (which P1 associated with anarchism) tended to use the same tools, creating a sense of \textit{``fear about these multiple perceptions''}. P14 recounted an incident involving a prominent left-wing community centre and action house which, unbeknownst to the community for years, had been set up by an undercover police officer who had held a leadership position before being revealed. P14 noted that this had made people more reflective about the potential impact of mundane decisions (e.g.~secure messaging and communication choices) on activist communities. Overall, using secure applications and other \textit{``hyper-secure''} (P1) practices led to concerns about how these actions could be interpreted by other group members -- the \emph{peer-to-peer gaze}.

A second type of social gaze participants sought to mitigate was \emph{the public gaze}. Many participants stated that engagement in activism in the UK often occurred alongside other activities (e.g.~their job). P1 phrased this as: \textit{``Sunday, Monday, Tuesday are company days. These days are my normal life. Wednesday, Thursday and Friday are my more extraordinary days, days when I work on activism stuff.''} Participants noted that this separation was important because of the climate movement's reputation as \textit{``young radicals or old hippie leftists''} (P5).

The final type of social pressure influencing participants in this study was \emph{the police gaze}. How participants felt their technology use would be perceived by the police in the event of a device being confiscated or infiltrated was of paramount importance. For example, P7 explained that they felt uncomfortable using applications like Signal because it would indicate to the police that they were involved in criminal activity. In general, participants wanted to avoid being seen in a certain way by the police, either to maintain safety or to maintain a personal sense of identity: \emph{They really make you feel like you're doing something wrong just by being on Telegram and Signal''} (P7). P10 explained that when they were recently arrested, the lack of Signal on their phone helped them to demonstrate their non-affiliation to certain activist groups. They described how they could show a limited connection to their group because there was no incriminating evidence on their phone and were allowed to leave with their phone. Rather than increasing secrecy in the face of police scrutiny, P10 noted that some activists felt pressured to open themselves up to the police and demonstrate their innocence.

\section{Discussion}\label{sec:discussion}

On the one hand our study confirms prior findings in related work with activists. This includes, for example, the perception of Signal being secure as observed in~\cite{AlbrechtCollective2021}, the adoption of practical workarounds, including the manual deletion of messages and hiding of applications, to protect against surveillance~\cite{DaffallaDefensive2021}, identity-related risks experienced in certain activist environments among, for example, LGBTQIA+ people~\cite{LernerPrivacy2020,USENIX:MccJenTal23}, and group-based constructions of information security within activist groups as identified in~\cite{AlbrechtCollective2021,SanchesUnder2020}, to mention a few. However, as we show, even such findings are fraught with conflicts. For example, as exemplified by Group E (cf.~\Cref{sec:group-e}), the autonomy afforded to group members meant that the collective decision of everyone to move to Slack was not followed by everyone. Thus, while common communication was seen as necessary to achieve collectively-agreed goals, the incompatible standpoints of autonomy made this common communication technologically impossible (cf.~\Cref{sec:paradox-autonomy}). Our findings also show how deliberations over whether or not to use some technology, e.g.~Zoom to enable greater accessibility, were rooted in considerations of virtue. Technological knowledge gaps became a moral rather than security question (cf.~\Cref{sec:techbros}).

Uncovering these conflicts necessitates immersion in what the authors of~\cite{CHI:KKSC16} refer to as the ``undercurrents'' of social movements; that which does not immediately present itself through interviews alone (\Cref{sec:rw-security-activists}). Our study shows, for example, how digitally-inclined activists were viewed with suspicion and social frustration in some groups, leading to an uneven adoption of tools they vouched for, like Signal, despite their (perceived) security benefits (\Cref{sec:findings-power-dynamics}). Therefore, our work shows how technologies that foreground other values beyond security in their design will be more uniformly useful to \emph{some} activist groups. Our findings also highlight that even the well-documented public-private dilemma is transformed in this setting as distinct gazes confronted participants. The question of presentation of self was not merely one of public-secret, it was multifaceted and multi-directional (\Cref{sec:findings-social-gazes}). 

Collectively, our findings point to how many current design or evaluation practices do not capture the complexity of activist decision-making or legitimacy concerns. However, as we note in~\Cref{sec:related-work}, while some of our findings resemble those reported in prior work on activists' security needs as well as broader research on `at risk' communities, we caution against reading our findings as generalisable to such wider contexts or, indeed, other activist settings. Rather, we stress the need to consider each setting in its own right; to develop ground-up and in-depth understandings of what security means to the \emph{specific} groups under study and how they consider, discuss and make security decisions individually and/or collectively. This is fundamental to an ethnographic programme of study.

We now discuss how participants made decisions about their own security and that of their group(s) through \emph{mundane} choices and disagreements surrounding communication. We show how an ethnographic approach enabled us to study participants' security concerns in their environments, including at protest sites and in activist meetings. We conclude by discussing what \emph{designing for} means in light of our findings.

\subsection{Mundane}\label{sec:mundane}

Our study and its findings sit within a long and active line of research on information security in social contexts: messaging application choice; a topic that has been the focus of many prior works in both security and HCI scholarship, e.g.~\cite{AlbrechtCollective2021,CHI:KoNaCh20}. Similar to what is reported in prior works, when discussing information security, participants in our study argued over whether to use Signal, WhatsApp or Telegram, among other messaging applications. As others have noted (e.g.~\cite{DaffallaDefensive2021}), political and social contexts shape discussions on technology choice for activists. This was also the case in our study.

Within this landscape, our findings uncover a distinctly different aspect of mundanity. What our findings add to prior work is \emph{how} the question of messaging applications was discussed by participants. The question of secure messaging was loaded with conflicting virtues that shaped how security was considered and reasoned about among participants. This was seen in how participants chose to adopt a particular (security) technology or practices to not stand out among their peers and to avoid the judgement of others (\cref{sec:findings-social-gazes}). We observed how decisions about security -- both for the individual and for the group -- were shaped not simply by the perception of a technology's security or usability, but by intertwining group dynamics and conflicting activist personas, ideals of self and virtues. Proposals by group members who were considered to have security or technology knowledge were not always adopted by other group members, rather their motives were thought of as `paranoia', cf.~the `digitally inclined' in Group E in \cref{sec:group-e}. In contrast, while organisers with a significant social standing, garnered through non-security related knowledge, espoused the use of Telegram, this led others to install the application despite being reluctant (\cref{sec:prestige}).

Observing such tensions is significant to information-security scholarship. They demonstrate how a group's information-security stance was often grounded in considerations that had only a tenuous connection to security (or technology) -- something not often foregrounded to the same extent in prior work. Instead, they were often shaped by the (conflicting) virtues that were held in high regard among climate activists, reflecting political priorities. Indeed, our findings suggest that for many participants, security served as the \emph{means} for multiple other conflicts: the negotiation and construction of activist identities and social relations. These findings highlight a contribution of this work: security trainers and activists considered digitally inclined may not portray the \emph{genuine} security perspectives of many other activists. This complicates the task for information-security researchers -- both practitioners and academics -- who seek to design for activists. Our work reinforces that understanding activists' information security necessitates studying their social contexts and relations which shape their conceptions and practice of security, the so-called ``undercurrents''~\cite{CHI:KKSC16}. While this insight is expressed in prior works, we consider the detailed documentation of such dynamics in our work a \usenixbf{key contribution}: we show \emph{how} interviewing the `digitally inclined' may lead to misleading conclusions.

\subsection{Methodology}\label{sec:methodology}

Our findings are derived from a relatively small number of interviews and participant observation. They thus build on the foundations established in much qualitative security research also based on interviews (\cref{sec:rw-security-activists}). However, to understand what designing \emph{for activists} actually means, the adherence to an ethnographic approach was necessary.

Through ethnographic encounters, we were able to capture participants' declarative and non-declarative knowledge (\cref{sec:interviews}), while bringing to light and making sense of the otherwise invisible ``undercurrents''~\cite{CHI:KKSC16} that shaped decision-making for the participants in our study. The ethnographic approach allowed us to ask different types of questions, while engaging with participants over time and building trust. It gave us access to spaces and discussions where approaches to security were being formulated, discussed and decided in more organic ways than what is possible through interviews alone. It was through this access and these engagements with participants in different settings that the underlying conflicts that we report in~\cref{sec:findings} were made visible. For example, by being present at group meetings, we observed how participants who showed a keen interest in technology were considered as suspicious and sometimes as undercover police (\Cref{sec:findings-social-gazes}). This assumption was apparent during action planning and during group-chat formation and membership management. We also observed during protests how security was considered particularly important to activists during and immediately before a protest, when the risk of police infiltration -- and thus disruption -- was considered particularly high (\Cref{sec:adversaries}). Thus, our presence \emph{in the field} highlighted the significance of time and place to the development and implementation of information protection practices. This also allowed us to observe contradictions and tensions between what was reported in interviews or agreed upon during group meetings, on the one hand, and how people actually approached and practised security either as part of their daily lives or at protests, on the other. This was, for example, observed in how mobile phones were usually carried during protests because important information was constantly being exchanged on a variety of applications and in different groups -- despite the fact that participants had reported and agreed that phones were left at home (\Cref{sec:group-e}). 

Our ethnographic grounding also meant that our study was indebted to reflexivity through an iterative process of data collection and analysis~\cite{Emerson11}. Indeed, it was through this analytical process involving the entire research team that we uncovered how central activist virtues were to participants' practice of information security. Reflexivity allowed us to situate the considerations about security and technology in the social worlds of the groups under study, which our work -- and that of others~\cite{DaffallaDefensive2021,AlbrechtCollective2021,ErmoshinaConcealing2022} -- has shown to be of critical import in activist settings. Staying with the ethnographic also means avoiding claims about \emph{saturation}, \emph{inter-rater reliability} and \emph{generalisability}, which appeal to a more neo-positivist rather than interpretative approach (e.g~\cite{QRSEH:BraCla21} and ethnographic writings such as~\cite{Emerson11,BrewerEthnographyBook2000}). 

The utility of ethnography lies in the fieldworker's immersion in the settings under study to bring to light and analyse insights that are often inaccessible through other methods. Thus, while we conducted formal interviews with 15 participants, it is through our ethnographic positioning that we were able to contextualise insights from what might otherwise be considered a limited number of interviews with the more subtle, often unspoken, observations during field encounters. In other words, ethnography captures the particularities -- the micro -- that less granular methods are not designed to capture.

Methodologically, our work is therefore a reflection on interview studies in/for security research. It highlights \emph{how} their limitations (which are well documented in prior works, cf.~\cref{sec:rw-security-activists}) may shape our findings. Given the essential role that interviews played in our study, a methodological \usenixbf{key contribution} of our work is a suggestion to reconsider how we \emph{conceptualise} interview studies in information-security research. We consider this category too broad and call for making more explicit that the \emph{style} of interview and the \emph{type} of analysis have implications for the \emph{type} of findings that is made possible.

\subsection{What Designing \emph{for Activists} Means}\label{sec:designing-for-activists}

Our findings speak also to the established programme in information-security research of designing for activists. Specifically, a \usenixbf{key contribution} based on our findings is to make explicit how \emph{designing for activists} can be a distinct activity to \emph{designing for activist settings}. The latter entails developing protocols, applications and security advice that speak to these settings. As discussed in~\cref{sec:rw-design-for-activists}, such design takeaways are common. The former programmes requires not only an understanding of the threat landscape facing activists, its temporal and spacial particularities, but also requires understanding the social structures and relations underpinning activism. It requires understanding the reasons activists have for their actions. These reasons may be informed by the self-reported or observed threat landscape but fundamentally use this threat landscape as the material to form a decision. There is no necessary transition from threat landscape to decision making. 

For example, a point of contention discussed by the participants in our study was whether to rely on Zoom for online meetings due to its reported collaboration with law enforcement. Yet, this contention was not one of operational security reflecting the threat landscape, but a moral one. The ideal activist self of some group members demanded to use Zoom for inclusivity reasons, and the ideal activist self of others demanded the abandonment of Zoom to protest its collaboration with law enforcement. As exemplified here -- and throughout our findings -- our work demonstrates how fundamental security decisions in some UK climate activist groups are made based on criteria far beyond technological considerations.
This necessitates understanding \emph{why} and \emph{how} decisions about application choices are made, not only the outcomes of such decisions. This also implies that insights for design take-aways may generalise more poorly than anticipated.

\subsection{Future Work}

Since our findings highlight how the specific context we studied shapes (in)securities and perceptions thereof, and also due to our exploratory research design and its inherent limitations (see~\Cref{sec:limitations}), this work does not give design recommendations. Its utility lies in bringing to light and examining otherwise hidden or underexplored particularities of how security is considered, discussed and decided upon within specific activist groups as part of the UK climate movement. Thus, looking ahead, our study suggests that future work should consider extended observational research in more diverse settings. Longer periods of embedded observation and engagements within such settings allow for findings that speak to how security dynamics change over time and are shaped by evolving threats and fears, changing group membership and internal conflicts. Such aspects are integral to understanding how new (security) technologies are adopted and rejected. Such a direction in future work has the potential to enable more meaningful participatory design efforts, deeply grounded in the settings under study.

\newpage

\section{Ethical Considerations}\label{sec:ethical-considerations}

Ethical approval for all research components was obtained from the fieldworker's institutional Research Ethics Committee (REC) prior to the start of the research. The research was classified as \emph{high risk} and a full risk assessment was carried out and approved by the Health \& Safety (H\&S) office of the fieldworker's institution. This included consideration of the risks of conducting \emph{in-situ} research during \emph{live} protests, with particular attention paid to the risk of confrontation between protesters and police. The research team worked with the institutional H\&S officers to implement safety protocols and risk mitigation measures, taking into account the recent legal developments in the UK (\cref{sec:rw-uk-climate-activism}). Before giving informed consent to take part, participants were provided with an information sheet (shared a few days in advance) and had the opportunity to ask questions. Participants had to be at least 18 years old to take part.

During participant observation, participants were informed of the reasons for the fieldworker's presence. In each setting, it was made clear that anyone's contributions would not be attributed to them or to a particular group, and risks of participation were enumerated. Participants were also asked if they were comfortable with the researcher's presence (everyone was). If anyone had objected to the fieldworker's presence at an event, the observation would not have commenced. Anyone present was also informed that they could speak with the fieldworker privately afterwards if they changed their mind about their participation. In protests, the fieldworker followed established practices for ethnographic observation in public spaces~\cite{CUNYEthnographyResearchGuide2020}, only noting non-identifying information in the field notes. Participants did not receive financial compensation, as we did not want participants to feel compelled to take part because of a financial incentive, and thus possibly accept increased risks. We do not include participant demographic summaries to minimise the risk of de-anonymisation, given the potential to link quotes with individuals.

Moreover, to protect the data, our analysis was conducted using Taguette~\cite{RampinTaguette2021}, an open source qualitative data analysis tool (similar to Lumivero's NVivo~\cite{NVivo}) that operates locally in an offline virtual environment.

\section{Open Science}\label{sec:open-science}
For safety reasons, we do not identify individual groups in this study and we refrain from detailing each group present during observations. Moreover, given the sensitive nature of this work, we have chosen not to share the full interview transcripts, interview scripts or field notes as part of our dataset. This decision is in line with established data protection practices in qualitative research, especially in contexts where sharing such data could put participants at (greater) risk and cause harm to them and/or their environments. The information shared during the fieldwork and interviews as well as the observations recorded in field notes, includes participants' individual and collective experiences, views and discussions. These cannot be abstracted to the extent that individuals may not be identified from the data -- even if specific identifiers are removed from the dataset. Further, building rapport with the groups under study is paramount in ethnographic research, which relies on mutual trust, often established over time and rooted in confidentiality. To adhere to ethical and responsible research practices, this must be respected at all stages of the research cycle, not least at the data dissemination stage. Rather than provide access to the raw dataset, we use pseudonymised and carefully curated quotes to support our findings and to ensure that the participants' voices remain present in our work. We make our interview guide available as an appendix.

\anonymous{}{
  \section*{Acknowledgements}

  This work would not have been possible without the generosity of participants who shared their time and trust with us, and the support of those who encouraged and enabled these conversations. We thank the reviewers and shepherd of this work for their thoughtful feedback. The research of Brough was supported by the EPSRC as part of the Centre for Doctoral Training in Cyber Security for the Everyday at Royal Holloway, University of London (EP/S021817/1).
}

\bibliographystyle{plain}
\bibliography{local}

\appendix

\section{Background on Technology}\label{sec:preliminaries-tech}

Discussions of security practises in our study often gravitated towards a choice of certain messaging applications.  We therefore here introduce the applications that were referenced in interviews and observed in the field.
\begin{description}[leftmargin=*]
\item[Signal] is an end-to-end encrypted protocol and application where users sign up with a phone number and user handles are either phone numbers or usernames. It supports one-on-one chat, group chats and (video) calls. Signal is often referred to as the gold standard of secure messaging both by practitioners and cryptographers. It has received several in depths analyses, e.g.~\cite{JC:CCDGS20}.
\item[WhatsApp] is the most popular messenger in the world with two billion users. It adopted an implementation of the Signal protocol in 2016, user handles are phone numbers and it supports end-to-end encrypted one-on-one chat, group chats and (video) calls. WhatsApp is owned by Meta\@. WhatsApp's encryption received a first (partial) security analysis in~\cite{AC:BalColGaj23}.
\item[Telegram] is a chat platform that supports one-to-one and large group messaging (up to 200,000 people) across multiple devices. End-to-end encryption is not enabled by default and must be invoked in secret chat and works only one-to-one mode. While phone numbers are required to sign up, these are not used as user handles. Telegram uses its own bespoke MTProto protocol both for transport (instead of TLS) and for end-to-end encryption. A first (partial) formal security analysis and some cryptographic attacks were given in~\cite{SP:AMPS22}.
\item[Threema] is a chat platform using the Ibex protocol. Threema uses only a Threema ID that is independent from email and phone number. It supports end-to-end encrypted chat, group chat and calls. Threema was studied in~\cite{USENIX:PatScaTru23} where several cryptographic attacks were presented.
\item[Session] is an end-to-end encrypted instant messaging application, supporting one-to-one and group chats. It does not require a phone number or email address to sign up (user handles are generated instead). It emphasises metadata avoidance and anonymity. No formal security analysis is available.
\item[Slack] is a messaging and VoIP platform that allows communication in one-to-one or in chat rooms or channels. It is commonly used in corporate environments and offers no end-to-end security, but only standard TLS-based transport security.
\item[Action Network] is an online organising platform that allows users to publish actions, fundraising forms, surveys, ticketed events, and letter writing campaigns. It also has tools for mass emailing and texting and all activists must sign up with their email address and phone number.
\item[CiviCRM] is an internal relationship management tool for non-profits and advocacy groups. Donors, members, event registrants, subscribers, grant-application seekers and funders, case contacts, group members, and volunteers can be managed.
\end{description}

\section{Interview Guide}\label{app:appendix-a}

\noindent The following is a list of interview questions used as a guide in this study. These questions were not part of a structured questionnaire; instead, they were asked flexibly to allow for a natural flow of conversation and to adapt to the context of each interview. As a result, the questions were not strictly adhered to, as per a semi-structured approach, and additional or follow-up questions were introduced based on participant responses. \\

\noindent \textbf{Introduction:}

\begin{enumerate}[leftmargin=*]
    \item \textbf{Explain the overall motivation for the project:}
    \begin{itemize}[leftmargin=0pt]
        \item The overarching goal of this study is to examine the security practices and perceptions of UK climate activists. We aim to examine the how climate activists think about security within their group, movement, and society more broadly.
        \item Answer questions about the purpose of this, why these sites, etc.
        \item Discuss their interest in the study and if they feel we can provide anything for them in terms of knowledge sharing. Clarify that no financial compensation will be provided for participation.
    \end{itemize}
    \item \textbf{Explain the purpose of this interview:}
	\begin{itemize}[leftmargin=0pt]
		\item  This project is being conducted in two ways -- 1) interviews (why we are here today) 2) observation of key activist spaces, such as meetings, trainings, and demonstrations.
		\item The goal of interviews is to understand the context of your activism, some general digital security concerns you have and your experiences more broadly.
	\end{itemize}
    \item \textbf{Obtain informed consent:}
    \begin{itemize}[leftmargin=0pt]
    \item Obtain consent in writing after confirming that they have read the Participant Information Sheet (PIS).
    \item Ask the participant if they wish to be recorded.
    \item Cover strategies to protect participants’ confidentiality such as eliminating all personally identifiable information, not using names, and storing all data securely.
    \item Answer any other questions they may have.
    \end{itemize}
\end{enumerate}

\noindent \textbf{Interview:}

\noindent Preliminaries: The participant describes themselves and their involvement with their group
\begin{enumerate}[leftmargin=*]
    \item What is the main goal of your group? Do you have a specific area or niche within the climate movement?
    \item What methods does your group employ to raise awareness about your cause?
    \item How are you funded?
    \item Is your group focused on achieving specific milestones or broader systemic changes?
\end{enumerate}

\noindent Ideas about threat / security: Participant describes broad conceptions of threat in terms of the climate movement
\begin{enumerate}[leftmargin=*]
    \item Have you faced resistance or opposition in your advocacy efforts? Can you give me an example?
    \item What are your main security concerns in your activism? What do you worry most about?
    \item What are some of the main threats to the type of work that you do?
\end{enumerate}

\noindent Relationship to technology: Participant describes how they generally feel about technology
\begin{enumerate}[leftmargin=*]
    \item What are your main priorities in terms of technology?
    \item How is tech useful for activism? Can you give me an example?
    \item How might tech be dangerous for activists? Can you give me an example?
    \item Does your group have any particular stance on technology?
    \item Do you make your own decisions about the type of tech that you use or do you follow the protocol or advice of others? E.g. people in your group, friends, etc?
\end{enumerate}

\noindent Ideas about digital security: The participant discusses cybersecurity
\begin{enumerate}
    \item Do you have any examples of what you consider secure technology and why?
    \item Do you ever worry about the security of information in terms of how your group organises and communicates?
    \item Does your group use any security technology for communication?
    \begin{itemize}
        \item If so, why did you chose the ones that you use?
        \item If not, why not?
    \end{itemize}
    \item What are some secure features that you find most useful for your activism?
    \item Who’s responsible to make decisions about security technology? (Does it even matter?)
    \item How do you feel about the types of messengers and communication channels that you use? Are there any things about them that bother you? Examples?
\end{enumerate}
\end{document}